\documentclass[twocolumn,epjc3]{svjour3}
\smartqed  
\RequirePackage{graphicx}
\usepackage{amsmath,amssymb,epstopdf}
\usepackage{color}

\journalname{Eur. Phys. J. C}

\begin{document}

\title{Quark matter supported wormhole in third order Lovelock gravity}


\author{Koushik Chakraborty\thanksref{e1,addr1}
        \and
        Abdul Aziz\thanksref{e2, addr2}
        \and
        Farook Rahaman\thanksref{e3,addr3}
        \and
        Saibal Ray\thanksref{e4,addr2}
        }

\thankstext{e1}{e-mail: koushik@associates.iucaa.in}
\thankstext{e2}{e-mail: azizmail2012@gmail.com}
\thankstext{e3}{e-mail: rahaman@associates.iucaa.in}
\thankstext{e4}{e-mail: saibal@associates.iucaa.in}


\institute{Department of Physics, Government College of Education, Burdwan 713102, West Bengal, India\label{addr1}
           \and
           Department of Physics, Government College of Engineering and Ceramic Technology, Kolkata 700010, West Bengal, India\label{addr2}
           \and
           Department of Mathematics, Jadavpur University, Kolkata 700 032, West Bengal, India \label{addr3}
           }

\date{Received: date / Accepted: date}

\maketitle

\begin{abstract}
It is generally believed that wormholes are supported by exotic matter violating Null Energy Condition (NEC). However, various studies of wormhole geometries under Lovelock theories of gravity have reported existence of wormhole supported by matter satisfying NEC. Being inspired by these results, we explore the possibility of the existence of wormhole supported by normal quark matter in third order Lovelock gravity theory. Well known MIT Bag Model Equation of state is chosen for describing the quark matter. Taking physically acceptable approximations, we solve the field equations for shape function which satisfies flare out condition. The residual of the approximate solution is studied for accuracy and found to be acceptable.
\end{abstract}



\section{Introduction}

Wormholes (WH) are tunnels in the spacetime topology connecting different regions of spacetime. Einstein Rosen bridge proposed by Einstein and Rosen \cite{ER1935} was first scientific work on WH solution in the framework of General Relativity. Einstein Rosen bridge was, however, not traversable and as a consequence, was not taken seriously by the scientific community. The term wormhole was coined by \cite{MW1957} and  Morris and Thorne \cite{MT1988} confirmed the physical acceptability of traversable WH and regenerated interests in the field in 1988. They proposed the traversable WH that satisfies the following:
\begin{itemize}
    \item the spacetime geometry must have no event horizon
    \item the matter sustaining the WH must violate classical NEC.
\end{itemize}
For a review on wormhole geometries one may refer \cite{visser1,lobo1}.

Construction of WH minimizing the support of exotic matter is an active field of research for a long time. This end, was achieved by several authors using one or other modified theories of gravity. Investigations based on Brans-Dicke theory ~[\cite{Agnese95}, \cite{Eiroa08} \cite{Anchordoqui97}, \cite{Nandi98}, \cite{He02}], Kaluza-Klein theory \cite{Shen91}, f($R$) gravity theory [\cite{Lobo09}, \cite{Garcia10}, \cite{Garcia11}, \cite{Harko13}, \cite{Taser16}], f($T$) theory \cite{Bohmer12}, f($G$) gravity \cite{Sharif15}, f($T, T_{G}$) gravity \cite{Sharif18} were reported. Rosa et al \cite{Rosa18} explored WH solution under a generalized hybrid metric-Palatini matter theory. They showed that matter field satisfies Null Energy Condition from the throat region upto infinity requiring no exotic matter.

Studying WH in Lovelock gravity theories is also an active field of research [\cite{shang}, \cite{bhawal}, \cite{tanwi}, \cite{dehghani}, \cite{mehdizadeh2012}, \cite{mehdizadeh2015}, \cite{zangeneh}, \cite{mehdizadeh2016}]. For some values of parameters, Richarte et al.  \cite{Richarte07} found that WH solution may exist supported by normal matter under Einstein-Gauss-Bonnet gravity theory.  Bandyopadhyay and Chakraborty~\cite{tanwi} reported that for suitable choices of parameter values, thin shell WH can be constructed of ordinary matter under Einstein-Yang-Mills-Gauss-Bonnet gravity. Lorentzian WH geometry supported by normal matter was explored by Dehghani and Dayyani~\cite{dehghani} for different choices of shape function and Lovelock coefficients. They reported the radius of shape function to be larger in third order Lovelock gravity compared to the Gauss Bonnet WHs. Whereas, in a study of WH solutions ~\cite{mehdizadeh2012} under higher dimensional Lovelock gravity theory, it was found that extra dimensions could support WHs.  Thin shell WHs were constructed unber third order Lovelock gravity by Mehdizadeh, Zangeneh and Lobo~\cite{mehdizadeh2015}. For some values of second and third order Lovelock coefficients, they found WHs supported by normal matter. Conditions for traversable WH solutions sustained by normal matter were explored in ~\cite{zangeneh} under third order Lovelock gravity with a cosmological constant term. Interestingly, the authors pointed out that a WH solution supported by normal matter corresponds to negative cosmological constant.Recently, Mehedizadeh and Lobo~\cite{mehdizadeh2016} also obtained exact WH solutions under third order Lovelock theory satsfying the energy conditions.

In a different context it is to note that, Rahaman et al.~\cite{rahaman2014a} considering the Navarro-Frenk-White (NFW)~\cite{NFW1996,NFW1997} density profile studied the possible existence of WH spacetime in the outer regions of galactic halo under GTR. Kuhfittig~\cite{kuhfittig} also reported a similar observation. In another work, Rahaman et al.~\cite{rahaman2014b} by employing the Universal Rotation Curve (URC)~\cite{URC2012} based dark matter model in the central part of the galactic halo and obtained analogous results. They also generalized the result to predict possible existence of wormholes in most of the spiral galaxies. On the other hand, Rahaman et al.~\cite{rahaman2016a,rahaman2016b} exploiting the NFW density profile as well as the URC calculated the tangential velocities $v^{\phi}$ of the test particles in the galactic halo under wormhole like line element. The result was satisfactory for the theoretical and observational plot within the range $9~kpc \leq r \leq 100~kpc$.

So, there are plethora of research articles reporting possibility of WH solutions supported by normal matter. In the present paper we investigate the possibility of existence of WH geometry supported by quark matter in third order Lovelock gravity. Strange quark matter, made up of up,  down and strange quarks, was shown to be energetically most favorable state of matter \cite{Witten84}. The core of a massive compact star is largely believed to be constituted of such matter. Researchers also predicted that massive compact stars might wholly be constituted of strange quark matter. The present study may shed light over the possibility of existence of WH geometry inside massive compact stars like hybrid stars, quark stars, strange stars etc. In the Section 2 we  present a brief outline of the Lovelock gravity theories. Section 3 presents the basic equations. The results and plots are discussed in Section 4. We conclude in the last Section 5.

\section{Brief outline of Lovelock gravity theory}\label{SecII}
In the framework of third-order Lovelock gravity, the action is 
\begin{equation}
I=\int d^{n+1}x\sqrt{-g}\left( L_{1}+\alpha _{2}^{\prime }L_{2}+\alpha _{3}^{\prime }L_{3}\right),  \label{love-act}
\end{equation}
assuming $8 \pi G_{n} =1$. Here, $G_{n}$ being the $n$-dimensional gravitational constant. $\alpha_{2}^{\prime}$ and $\alpha_{3}^{\prime}$ are the second (Gauss-Bonnet) and third order Lovelock coefficients, $g$ is the determinant of the metric, $L_{1}=R$ is the Einstein-Hilbert Lagrangian, the term $L_{2}$ is the Gauss-Bonnet Lagrangian given by
\begin{equation}
L_{2}=R_{abcd}R^{abcd}-4R_{ab}R^{ab}+R^{2},
\end{equation}
and the third order Lovelock Lagrangian $L_{3}$ is defined as
\begin{eqnarray}
L_{3} =&&2R^{abcd}R_{cdmn}R_{\phantom{mn}{ab}}^{mn}+8R_{\phantom{ab}{cm}}^{ab}R_{\phantom{cd}{bn}}^{cd}R_{\phantom{mn}{ad}}^{mn} \nonumber \\&&
    +24R^{abcd}R_{cdbm}R_{\phantom{m}{a}}^{m}+3RR^{abcd}R_{cdab}
    \notag \\
&&+24R^{abcd}R_{ca}R_{db}+16R^{ab}R_{bc}R_{\phantom{c}{a}}^{c} \nonumber \\&&
    -12RR^{ab}R_{ab}+R^{3}.
\end{eqnarray}

In Lovelock theory, for an $n$-dimensional space, only terms with order less than $\left[\frac{(n+1)}{2} \right]$ contribute to the field equations. Here, we used the notation that $\left[\frac{n}{2} \right]$ will give biggest integer less than $\frac{n}{2}$. Since we are considering third order Lovelock gravity, its effects will be apparent for $n \geq 7$.

Thus, Varying the action (\ref{love-act}) with respect to the metric we get the field equations up to third order as follows: 
\begin{equation}
G^{E}_{ab} + \alpha _{2}^{\prime } G^{(2)}_{ab} + \alpha _{3}^{\prime } G^{(3)}_{ab} = T_{ab},
\end{equation}
where $T_{ab}$ is the energy-momentum tensor, $G_{ab}^{E}$ is the Einstein tensor whereas $G_{ab}^{(2)}$ and $G_{ab}^{(3)}$ are given by

\begin{eqnarray*}
G_{ab}^{(2)} =&& -2R_{acdn}R_{\phantom{cnd}{b}}^{dnc}-4R_{ambc}R^{mc} \nonumber\\&&
-4R_{ac}R_{\phantom{c}b }^{c} 
    +2RR_{ab})-\frac{1}{2}L_{2}g_{ab} \,,
    \\
G_{ab}^{(3)} =&& -3(4R^{nmcd}R_{cdpm}R_{\phantom{p }{b n a}}^{p}-8R_{\phantom{nm}{pc}}^{nm}R_{\phantom{cd}{na}}^{cn}R_{\phantom{p}{bmd}}^{p }
	\nonumber\\
	&&+2R_{b}^{\phantom{b}{ncd}}R_{cdpm}R_{\phantom{pm}{na}}^{pm}-R^{nmcd}R_{cdnm}R_{ba}
 	\nonumber\\
	&&+8R_{\phantom{n}{jkm}}^{n}R_{\phantom{kl}{ni}}^{kl }R_{\phantom{m}l }^{m }+8R_{\phantom{k}{jnl}}^{k }R_{\phantom {nm}{ki}}^{nm }R_{\phantom{l}{m}}^{l }
	\nonumber\\
	&&+4R_{b }^{\phantom{b}{n c d}}R_{c da m }R_{\phantom{m}{n}}^{m }-4R_{b}^{\phantom{b}{n c d}}R_{c d n m}R_{\phantom{m}{a}}^{m } \notag\\&&
+4R^{n m c d}R_{c d n a }R_{b m}
    \nonumber\\&&
	+2RR_{b }^{\phantom{b}{d n m}}R_{n m d a } +8R_{\phantom{n}{b a m }}^{n }R_{\phantom{m}{c}}^{m }R_{\phantom{c}{n}}^{c}
	\nonumber\\
	&&-8R_{\phantom{c}{b nm }}^{c }R_{\phantom{n}{c}}^{n }R_{a }^{m }-8R_{\phantom{n }{c a}}^{n m }R_{\phantom{c}{n}}^{c}R_{b m }
	\nonumber\\
	&&-4RR_{\phantom{n}{b a m }}^{n }R_{\phantom{m}n }^{m }+4R^{n m }R_{m n }R_{ba}-8R_{\phantom{n}{b}}^{n}R_{n m }R_{\phantom{m}{a}}^{m }
	\nonumber\\
	&&+4RR_{b m }R_{\phantom{m}{a }}^{m }-R^{2}R_{b a})-\frac{1}{2}L_{3}g_{ab}.
\end{eqnarray*}

\section{Basic Equations}\label{SecIII}
The $n$-dimensional traversable wormhole metric is given by
\begin{equation}\label{E:line1}
ds^2=-e^{2 \Phi(r)}dt^2+\frac{dr^2}{\left(1-\frac{b(r)}{r}\right)}+r^2 d\Omega_{n-2}^2,
\end{equation}
using units in which $c = G = 1$. Here $\Phi(r)$ is the redshift function which must be everywhere finite to prevent the event horizon, $b(r)$ is the shape function and $d\Omega_{n-2}^2$ is the metric on the surface of a $(n-2)$-sphere. The shape function of the wormhole essentially satisfies the condition $b(r_0) = r_0$ at $r = r_0$ where $r_0$ is the throat of the wormhole. This condition is commonly known as the flare-out condition which gives at the throat $b'(r_0) < 1$ while $b(r) < r$ near the throat. In the present paper, for the sake of convenience we shall take the function $\Phi$ to be constant.

The energy-momentum tensor is given by
\begin{equation}
T^{\mu}_{\nu} = diag[ - \rho(r), p_{\|}(r), p_{\bot}(r), p_{\bot}(r), \dots],
\end{equation}
where $\rho(r)$ is the energy density, $p_{\|}(r)$ is the radial pressure and $p_{\bot}(r)$ gives the transverse pressure.

With the above assumption for the redshift function, $\Phi(r) = constant$, the Einstein equations for the above mentioned metric are as follows
\[ \rho(r) = -\frac{(n-2)}{2r^2}\left(1+\frac{2\alpha_2 b}{r^3}+\frac{3\alpha_3 b^2}{r^6}\right)\frac{(b-rb^{\prime})}{r}\]
\begin{equation} \label{rho}
+\frac{(n-2)b}{2r^3}\left[(n-3)+(n-5)\frac{\alpha_2 b}{r^3}+(n-7)\frac{\alpha_3 b^2}{r^6}\right],
\end{equation}
\[ p_{\|}(r) = -\frac{(n-2)(n-3)b}{2r^3} \]
\begin{equation} \label{pr}
 +(n-5)\frac{\alpha_2 (n-2)b^2}{2r^9}+(n-7)\frac{\alpha_3(n-2) b^3}{2r^9},
\end{equation}

\[p_{\bot}(r) = \Xi(r) \left[(n-3) + (n-5)\frac{2 \alpha_2 b}{r^3} + (n-7)\frac{3 \alpha_3 b^2}{r^6} \right]\]
\[ - \frac{b}{2 r^3}(n-3)(n -4) - (n-5)(n -6)\frac{\alpha_2 b^2}{2r^6} \]
\begin{equation} \label{pt}
  - (n-7)(n -8)\frac{\alpha_3 b^3}{2r^9},
\end{equation}
where $\Xi = \left(1 - \frac{b}{r} \right) \left(\frac{(b - rb')}{2r^2(r -b)} \right)$, $\alpha_2 = (n-3)(n-4) \alpha_{2}'$ and $\alpha_3 = (n-3)(n-4)(n-5)(n-6) \alpha_{3}'$. 

In the above equations the prime denotes the derivative with respect to $r$. Here, we obtain a system of three independent nonlinear equations \ref{rho}, \ref{pr}, \ref{pt}. We have to solve for four unknown functions, $\rho(r)$, $p_{\|}(r)$, $p_{\bot}(r)$ and $b(r)$. The redshift function, $\Phi(r)$ is already assumed to be constant implying zero tidal force. In order to close the system we assume a specific equation of state (EOS). We assume the existence MIT bag model EOS is given by 
\begin{equation}\label{bagEOS}
4B_g = \rho - 3p_r.
\end{equation}

\section{Results and Discussion}

\subsection{Shape function}
In simplified form Eq. \ref{rho} can be written as
\begin{equation}
\rho(r) = \frac{(n-2)}{2}\left[r \xi^{\prime}+(n-1)\xi \right],\label{rhosimp}
\end{equation}
where $\xi=\frac{b}{r^3}\left[1+\alpha_2 \frac{b}{r^3}+\alpha_3 \frac{b^2}{r^6}\right]$.

With the same considerations for simplification Eq.\ref{pr} reduces to
\begin{equation}
p_{\|} (r) = -\frac{(n-2)}{2} \left[(n-7) \xi + 2 \frac{b}{r^3} \left(2 + \alpha_2 \frac{b}{r^3}\right)  \right].\label{pr2}
\end{equation}

With the assumption of EOS given in Eqs. \ref{bagEOS},~\ref{rhosimp} and~\ref{pr2} reduces to the form
\begin{equation}
r \xi^{\prime} +  2(2n-11) \xi + 6 \frac{b}{r^3} \left(2 + \alpha_2 \frac{b}{r^3}\right) = \frac{8B_g}{n-2}. \label{final-eq}
\end{equation}

Now, without any loss of generality, we may assume the following mathematical relationships between the modified second and third order coupling constants of Lovelock action i.e., $\alpha_2 = (n-3)(n-4) \alpha_{2}'$ and $\alpha_3 = (n-3)(n-4)(n-5)(n-6) \alpha_{3}'$.

As it is hard to find exact analytical solution  non-linear  Eq. \ref{final-eq}. For the mathematical simplicity let us  first consider
\begin{equation}
\alpha_3 = \left(\frac{\alpha_2}{2}\right)^2, \label{cond1}
\end{equation}
then 
\begin{equation}\label{xi1}
\xi =\frac{b}{r^3}\left(1+\frac{\alpha_2}{2}\frac{b}{r^3}\right)^2.
\end{equation}

We now consider the coupling coefficient for the second order Lovelock coefficient to be sufficiently small such that $| \alpha_2 | b \ll r^3$ for $| \alpha_2 | < 1$. With this approximation, the above expression for $\xi$ in Eq. \ref{xi1} reduces to the following form
\begin{equation}
\xi =\frac{b}{r^3}\left(1+\alpha_2 \frac{b}{r^3}\right),
\end{equation}
i.e.,
\begin{equation}
\frac{b}{r^3} = \frac{1}{2\alpha_2}\left(-1+\sqrt{1+4\alpha_2 \xi}\right).
\end{equation}
 
For weak coupling, under similar arguments as before, Eq. \ref{final-eq} reduces to the form
\begin{equation}
r \xi^{\prime} +  l \xi  = C,
\end{equation}
where $C = \frac{8B_g}{n-2}$ and $l = 2(2n-5)$. The solution of this equation is
\begin{equation}
\xi  = \frac{C}{l}+ C_1 r^{-l}, \label{xi2}
\end{equation}
where $C_1$ is constant of integration. 

Hence the shape function is given by
\begin{equation}\label{b1}
b(r) = \frac{r^3}{2\alpha_2}\left[-1+\sqrt{1+4\alpha_2 \left(\frac{C}{l}+ C_1 r^{-l}\right)}\right].
\end{equation}

Note that in the limit $\alpha_2 \rightarrow 0$, we actually get the results for $4$ dimensional general relativity, since we have already assumed Eq. \ref{cond1}. Taking this limit on the Eq. \ref{b1}, we get
\begin{equation}
	b(r)|_{\alpha_2 \rightarrow 0}= r^3 \left[\frac{2B_g}{3}+\left(1-\frac{2B_g}{3}r_0^2\right)r^4_0 r^{-6}\right],
\end{equation}
which must be a decreasing function for $r_0 < 1$.

Taking the throat of the WH at $r = r_0$, from condition $b(r_0) = r_0$ gives 
\begin{equation}
C_1=\frac{l ({r_0}^2+ \alpha_2)-C {r_0}^4}{l{r_0}^{4-l}}.
\end{equation}

Now, we can check the acceptability of the approximation $| \alpha_2 | b \ll r^3$ for $| \alpha_2 | < 1$, taken in the calculation of $b(r)$. 

\begin{figure} [thbp]
	\includegraphics[width=6.0cm]{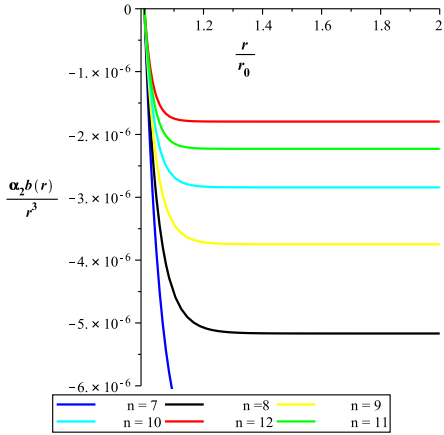}
	\caption { Plot showing the values of $\frac{| \alpha_2 | b}{r^3}$ for $n = 7, 8, 9, 10, 11$ for values $r_0=0.9$, $\alpha_2=-0.81$, $B_g=0.0001052631579$ km$^{-2}$.}\label{condn}
\end{figure}

As can be noted from Fig. \ref{condn}, the values of $\frac{| \alpha_2 | b}{r^3}$ are acceptably small everywhere near the throat of the wormhole. 
To test the accuracy for solution of shape function we define \cite{Finlayson66} residual of shape function as given by 
\[ Res[b(r)] = r \xi^{\prime} +  2(2n-11) \xi + \]
\begin{equation}
	 6 \frac{b}{r^3} \left(2 + \alpha_2 \frac{b}{r^3}\right) - \frac{8B_g}{n-2},
\end{equation}
where for exact solution Res[b(r)] would be zero. 

\begin{center}
\begin{figure*}[thbp]

		\includegraphics[width=6.0cm]{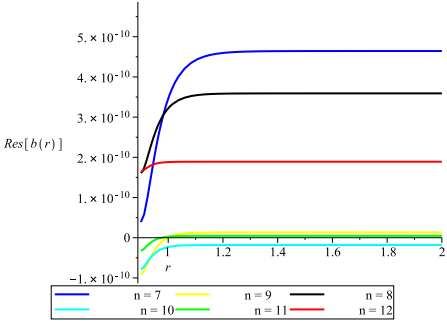}
	\caption{\small{Plot to study residual for shape function taking $r_0=0.9$, $\alpha_2=-0.81$, $B_g=0.0001052631579$ km$^{-2}$ for different value of $n$.} }\label{resb}
\end{figure*}
\end{center}

From Fig \ref{resb} note that Res[b(r)] is very small compared to shape function value which indicates the accuracy of  analytic solution under weak coupling approximation is very high.

\begin{figure*}[thbp]
	\begin{tabular}{lr}
		\includegraphics[width=6.0cm]{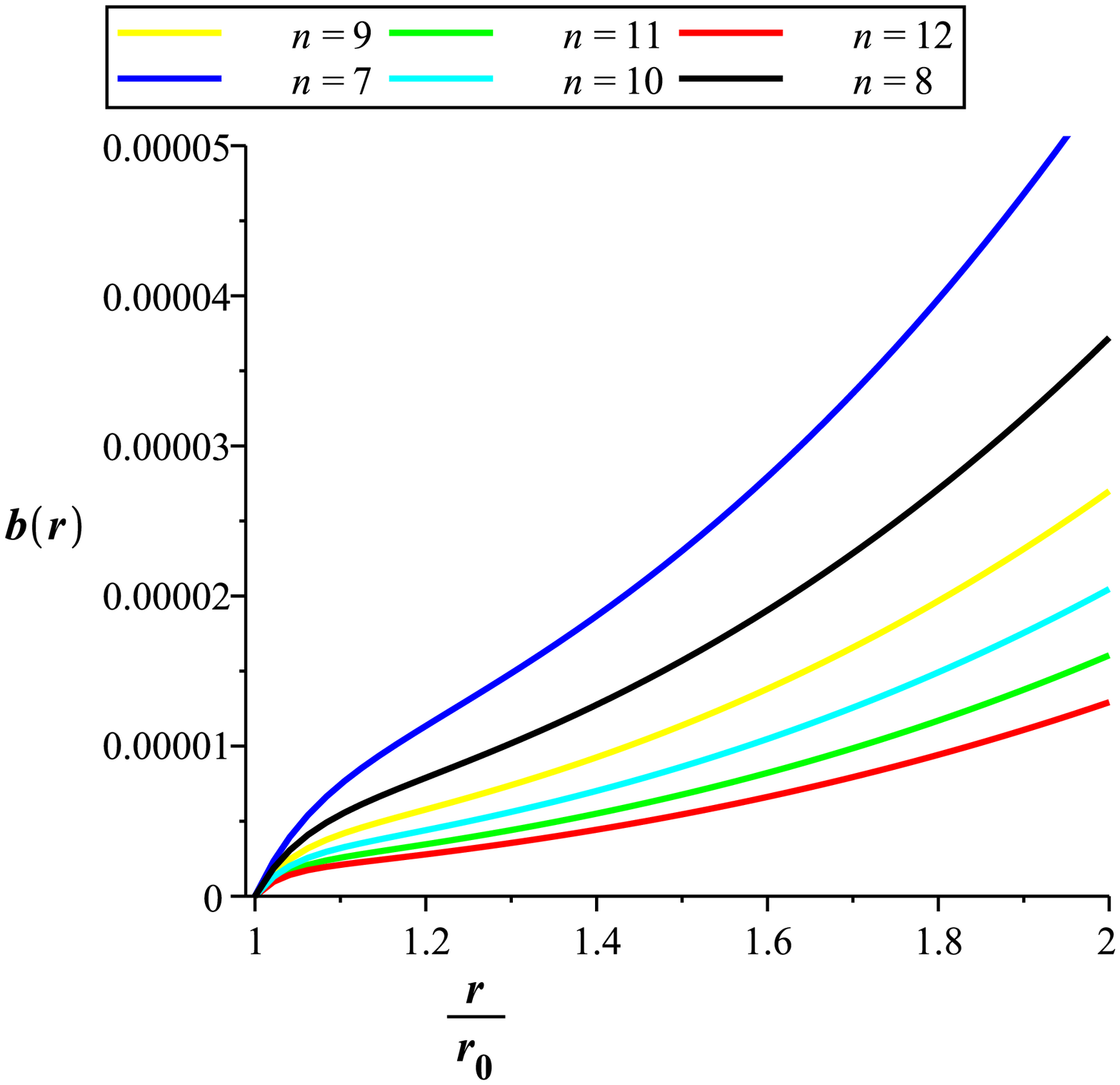}&
		\includegraphics[width=6.0cm]{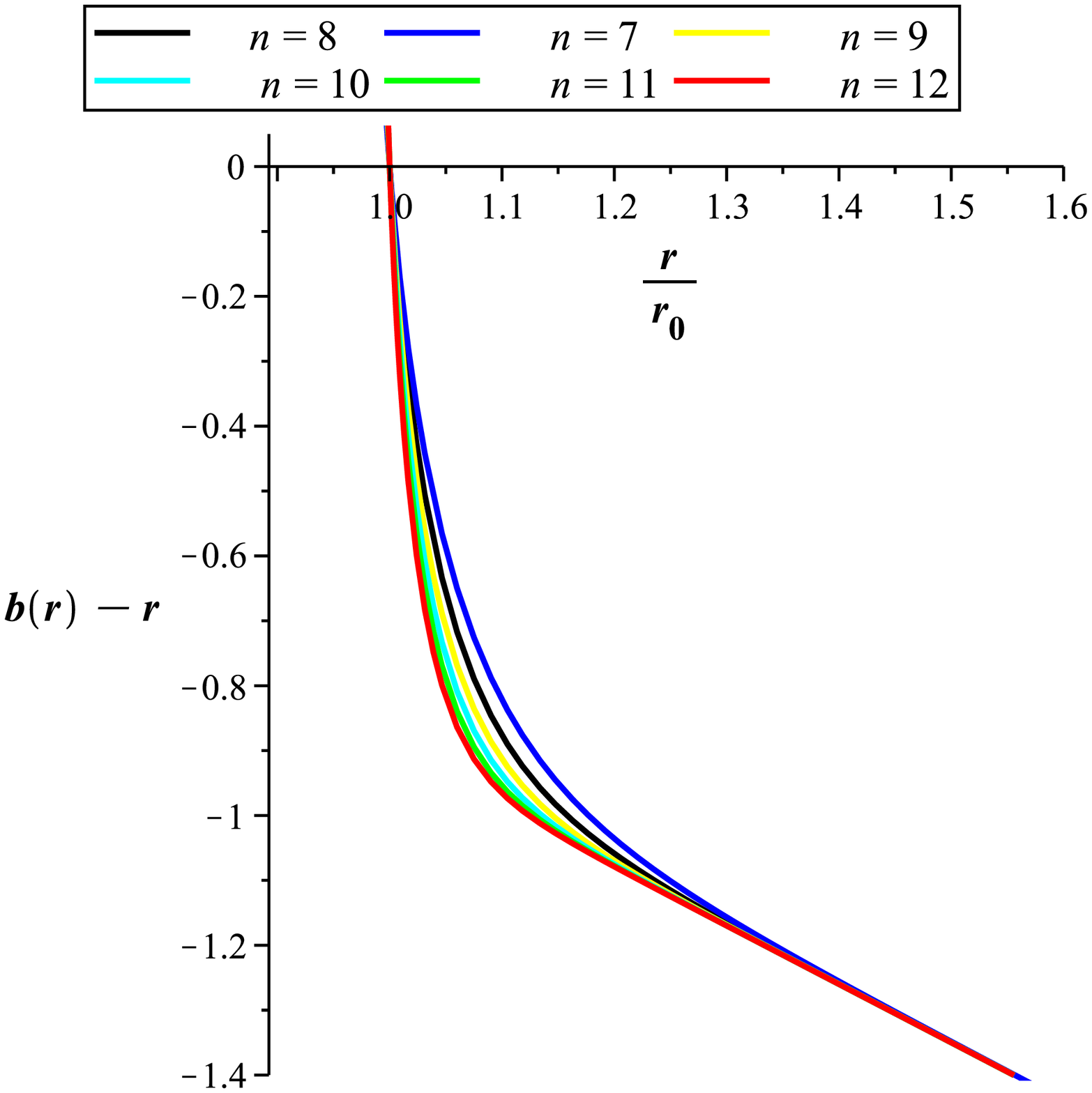}
		\end{tabular}
	\caption{\small{Plot to study $b(r)$ (left) and $b(r) - r$ (right) taking $r_0=0.9$, $\alpha_2=-0.81$, $B_g=0.0001052631579$ km$^{-2}$ for different value of $n$.} }\label{b(r)}
\end{figure*}

The plot in the left of Fig. \ref{b(r)} shows that shape function is positive in the entire region for $n = 7, 8, 9, 10, 11$ and it is increasing function as well. From the plot in right of Fig. \ref{b(r)}, it can be noted that $b(r)<r$ for $r>r_0$ which means flare out condition is satisfied for $n = 7, 8, 9, 10, 11$. 

\begin{figure}[thbp]
	\includegraphics[width=6.0cm]{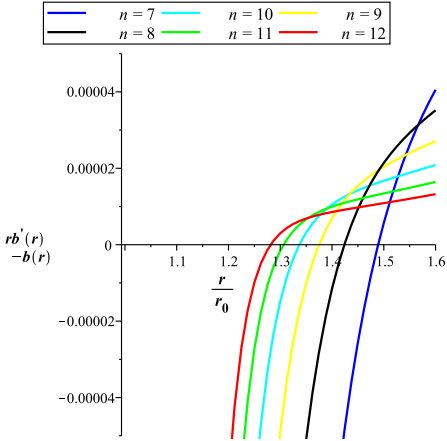}
	\caption {Flare out condtion of the wormhole taking $r_0=0.9$, $\alpha_2=-0.81$, $B_g=0.0001052631579$ km$^{-2}$ for different value of $n$.} \label{fc1}
\end{figure}

Yet another relation $rb^{\prime}(r) < b(r)$  is satisfied as shown in Fig. \ref{fc1}. The condition $rb^{\prime}-b=0$ gives the maximum physical radial distance ($r_{max}$) for WH.

\subsection{Density and the coupling coefficients}
Substituting Eq. \ref{xi2} in Eq. \ref{rhosimp} we get 
\[ 	\rho(r) =  \]
\begin{equation}
\frac{3(n-2)(n-3)(Cr_0^4-lr_0^2-l\alpha_2)}{4 (2n-5) r_0^{4-l}}r^{-l} +\frac{2B_g (n-1)}{(2n-5)}.
\end{equation}

For positive energy density we will have $n > 2$ and $(Cr_0^4-lr_0^2-l\alpha_2)>0$ which gives upper limit for $\alpha_2$ as given by 
\begin{equation}
	(\alpha_{2})_{max} = r_0^2 \left(\frac{4B_g }{(n-2)(2n-5)}r_0^2 -1 \right),  \label{alphamax}
\end{equation}
which is positive for $B_g=0.0001052631579$ km$^{-2}$ only when $r_0$ is the order of $10^2$ resulting large values of $(\alpha_{2})_{max}$. However, the present model is valid for only smaller values of $\mid (\alpha_{2})_{max} \mid<<1$ which is true for $r_0<1$ and negative $\alpha_2$ as indicated by Eq. \ref{alphamax}. $\alpha_2$ is negative considering positive energy density and validity of the present model.

Note that the present model of wormhole is physically valid for the range of parameter values  $0<r_0<1$ and  $-1<\alpha_2<0$. We have taken different data set for ($r_0$, $\alpha_2$) for analysis of the model.The graphical nature of the physical functions and conditions is shown taking data set ($r_0=0.9$, $\alpha_2=-0.81$, $B_g=0.0001052631579$ $km^{-2}$) for different value of $n$.

\begin{table}[thbp]
  \caption{Study of the effect of parameter values on the physical acceptance of the model} \label{tbl-1}
 	
 	\begin{tabular}{|c|c|c|c|c|c|c|c|c}
 		\hline
 		$r_0$  & $\alpha_2$   &  $b(r)$ & $b(r) - r$ & $r_{max}$ & $rb^{\prime}-b(r)$  \\
 		\hline
 		0.9  & -0.81 & +ve \& increasing  & -ve  for $r>r_0$ & $ 1.489 r_0$ & satisfied  \\
 		\hline
 		0.8 &  -0.64 & +ve \& increasing  & -ve  for $r>r_0$  & $ 1.529 r_0$  & satisfied   \\
 		\hline
 		0.7 & -0.49 & +ve \& increasing & -ve  for $r>r_0$ & $ 1.575 r_0$  & satisfied   \\
 		\hline
 		0.6 &  -0.36 & +ve \& increasing & -ve  for $r>r_0$ & $1.629 r_0$  & satisfied   \\
 		\hline
 		0.5 &  -0.25 & +ve \& increasing & -ve  for $r>r_0$ & $1.697 r_0$   & satisfied  \\
 		\hline
 		0.4  & -0.16 & +ve \& increasing & -ve  for $r>r_0$ & $1.784 r_0$  & satisfied  \\
 		\hline
 		0.3 & -0.09 & +ve \& increasing & -ve  for $r>r_0$ & $ 1.901 r_0$   & satisfied   \\
 		\hline
 		0.2 &  -0.04 & +ve \& increasing & -ve  for $r>r_0$ & $ 2.080 r_0$   & satisfied  \\
 		\hline
 		0.1 &  -0.01 & +ve \& increasing & -ve  for $r>r_0$ & $ 2.427 r_0$  & satisfied    \\
 		\hline
 		
 	\end{tabular}
 \end{table}

\begin{table}[h]
	\caption{Study of the effect of parameter values on the satisfaction of energy conditions for the WH} \label{tbl-2}
	
	\begin{tabular}{|c|c|c|c|c|c|c|c|c}
		\hline
		$r_0$  & $\alpha_2$ & $\rho$ \& $\rho+p_{\|}$ \& $\rho+p_{\bot}$ & $\rho+p_{\|}+ (n-2)p_{\bot}$   \\
		\hline
		0.9  & -0.81  & +ve \& decreasing  & -ve \& increasing   \\
		\hline
		0.8 &  -0.64 & +ve \& decreasing   & -ve \& increasing   \\
		\hline
		0.7 & -0.49 & +ve \& decreasing  & -ve \& increasing    \\
		\hline
		0.6 &  -0.36 & +ve \& decreasing  & -ve \& increasing    \\
		\hline
		0.5 &  -0.25 & +ve \& decreasing  & -ve \& increasing  \\
		\hline
		0.4  & -0.16 & +ve \& decreasing  & -ve \& increasing  \\
		\hline
		0.3 & -0.09 & +ve \& decreasing  & -ve \& increasing    \\
		\hline
		0.2 &  -0.04 & +ve \& decreasing  & -ve \& increasing   \\
		\hline
		0.1 &  -0.01 & +ve \& decreasing  & -ve \& increasing     \\
		\hline
		
	\end{tabular}
\end{table}

\subsection{Energy conditions}
The standard point-wise energy conditions of classical general relativity are helpful for extracting significant information of the matter distribution without assuming a particular equation of state. We consider three energy conditions, namely, Null Energy Condition (NEC), Weak Energy Condition (WEC), Strong Energy Condition (SEC).  

In terms of principal pressures the NEC is given by following inequations:
\begin{eqnarray}
\rho + p_{\|} \geq 0, 
\rho + p_{\bot} \geq 0, \label{NEC}
\end{eqnarray}
whereas WEC is given by following inequations:
\begin{eqnarray}
\rho \geq 0,\label{WEC1}
\rho + p_{\|} \geq 0,
\rho + p_{\bot} \geq 0.\label{WEC}
\end{eqnarray} 

The SEC can be expressed as:
\begin{eqnarray}
\rho + \Sigma_{i} p_{i} \geq 0,
\rho + p_{i} \geq 0. \label{SEC}
\end{eqnarray}

With the arguments of weak coupling as mentioned earlier, the radial pressure given in Eq. \ref{pr2} reduces to 
\begin{equation}\label{prsimp}
p_{\|}  (r) = -\frac{(n-2)(n-3)}{2} \xi. 
\end{equation}

For our present model of WH, to verify the NEC, WEC and SEC, we consider the following equations respectively:
\begin{equation}
\rho + p_{\|} = \frac{(n-2)}{2}\left(r \xi^{\prime}+2\xi \right),
\end{equation}
 
\begin{equation}
\rho + p_{\bot} =  r \xi^{\prime} + (n-2) \xi + 2 \alpha_2 r \xi \xi^{\prime},
\end{equation}
\[ \rho+p_{\|}+ (n- 2) p_{\bot}= \] 
\begin{equation}
 -\frac{(n-2)}{2} \left[ (n-4) \{ r \xi^{\prime} + (n - 1) \xi  \} - 4 \alpha_2 r \xi \xi^{\prime}\right].
\end{equation}

\begin{figure*}[thbp]
	\begin{tabular}{lr}
		\includegraphics[width=6.0cm]{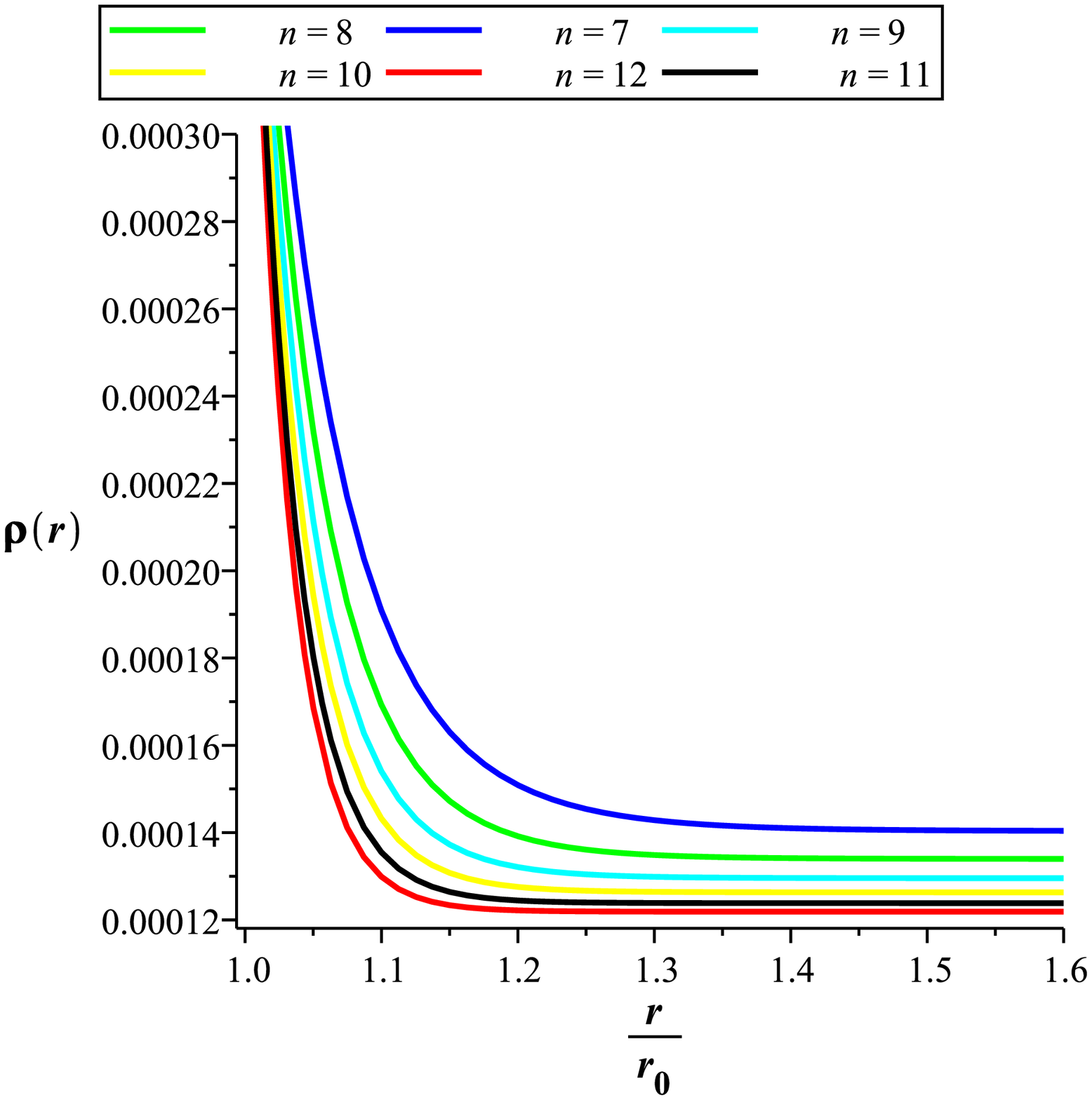}&
		\includegraphics[width=6.0cm]{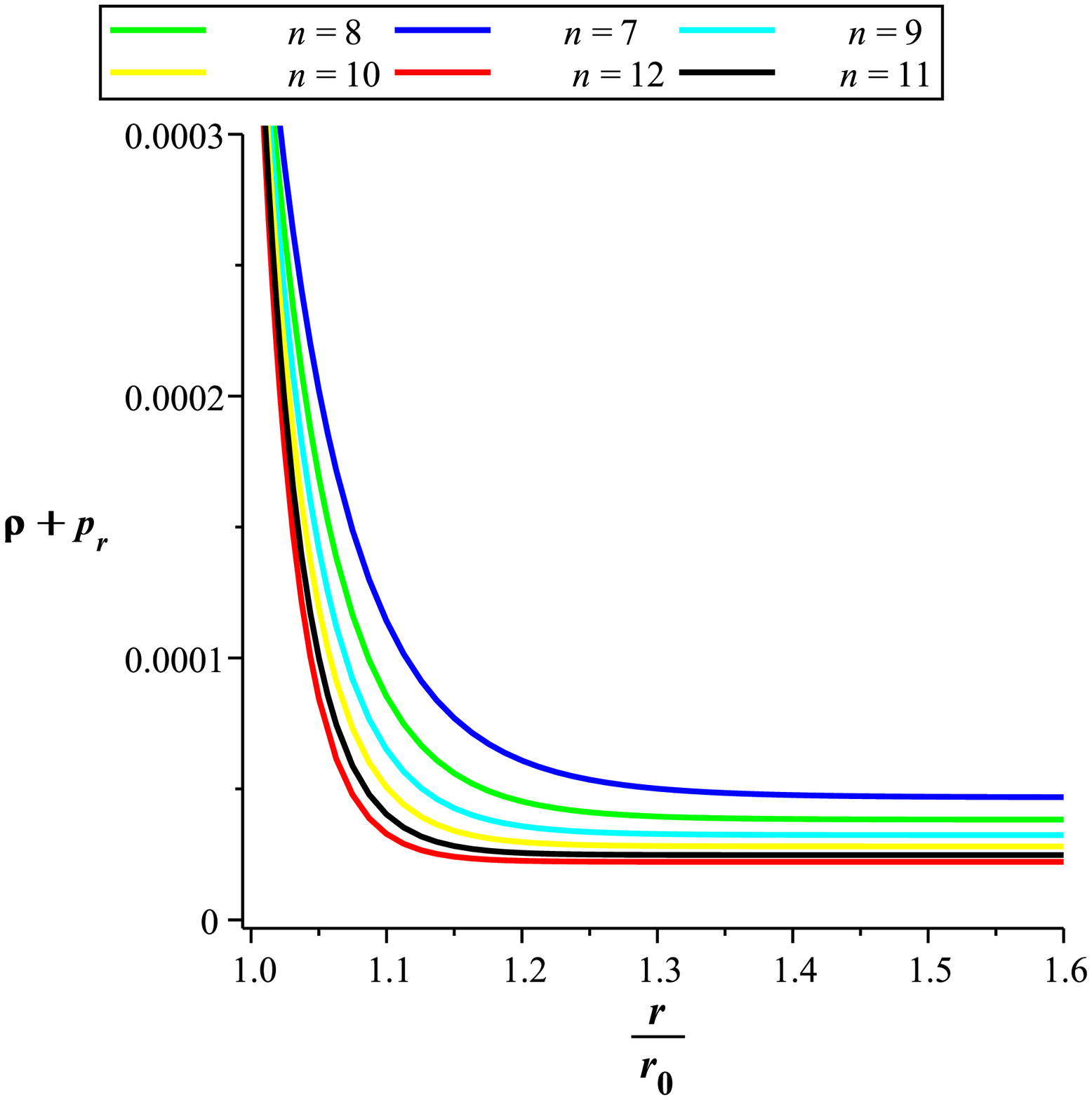}
	\end{tabular}
	\caption{\small{Plot to study $\rho$ and $\rho + p_{r}$ taking $r_0=0.9$, $\alpha_2=-0.81$, $B_g=0.0001052631579$ km$^{-2}$ for different value of $n$.} }\label{EC1}
\end{figure*}

\begin{figure*}[thbp]
	\begin{tabular}{lr}
		\includegraphics[width=6.0cm]{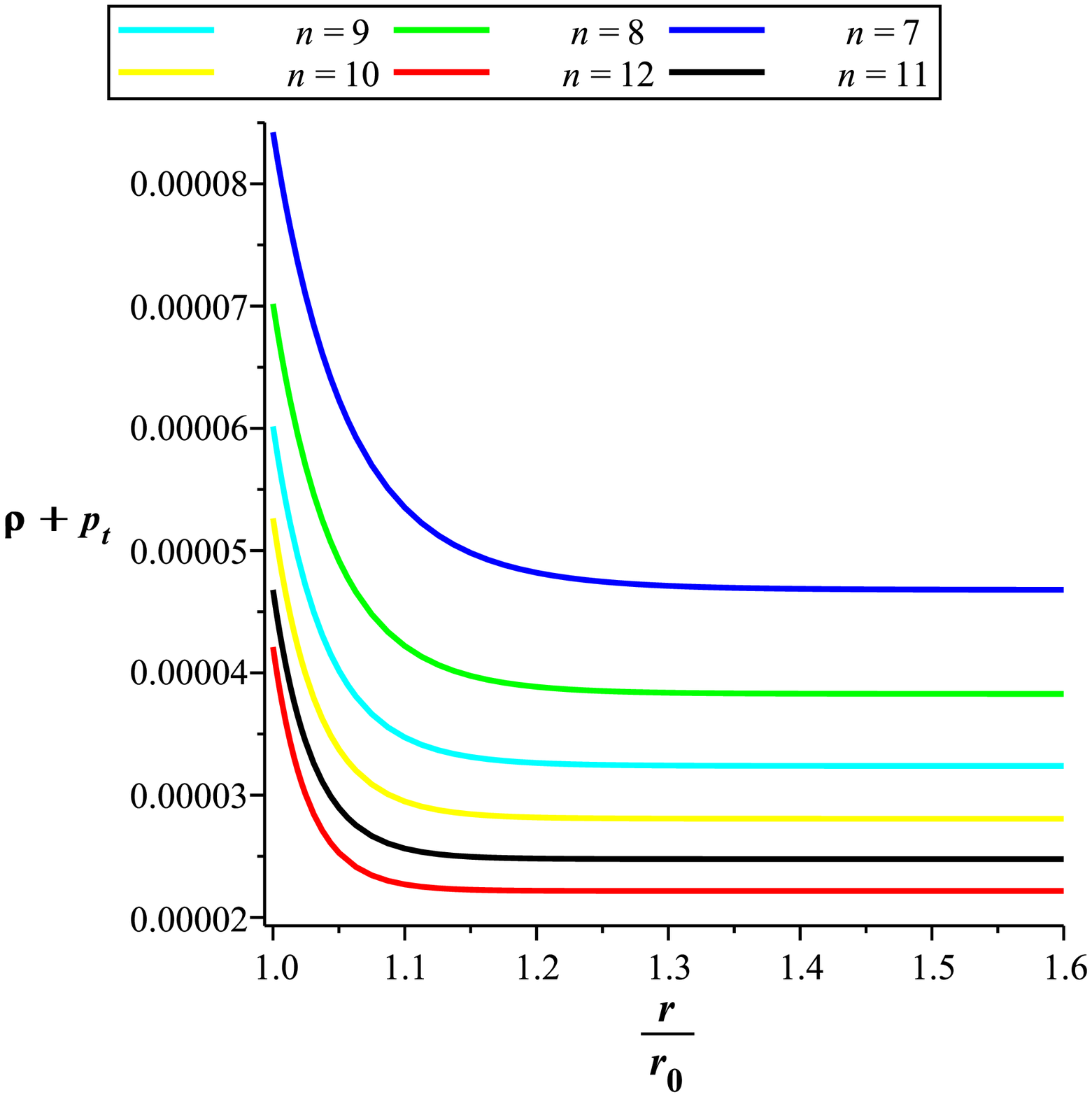}&
		\includegraphics[width=6.0cm]{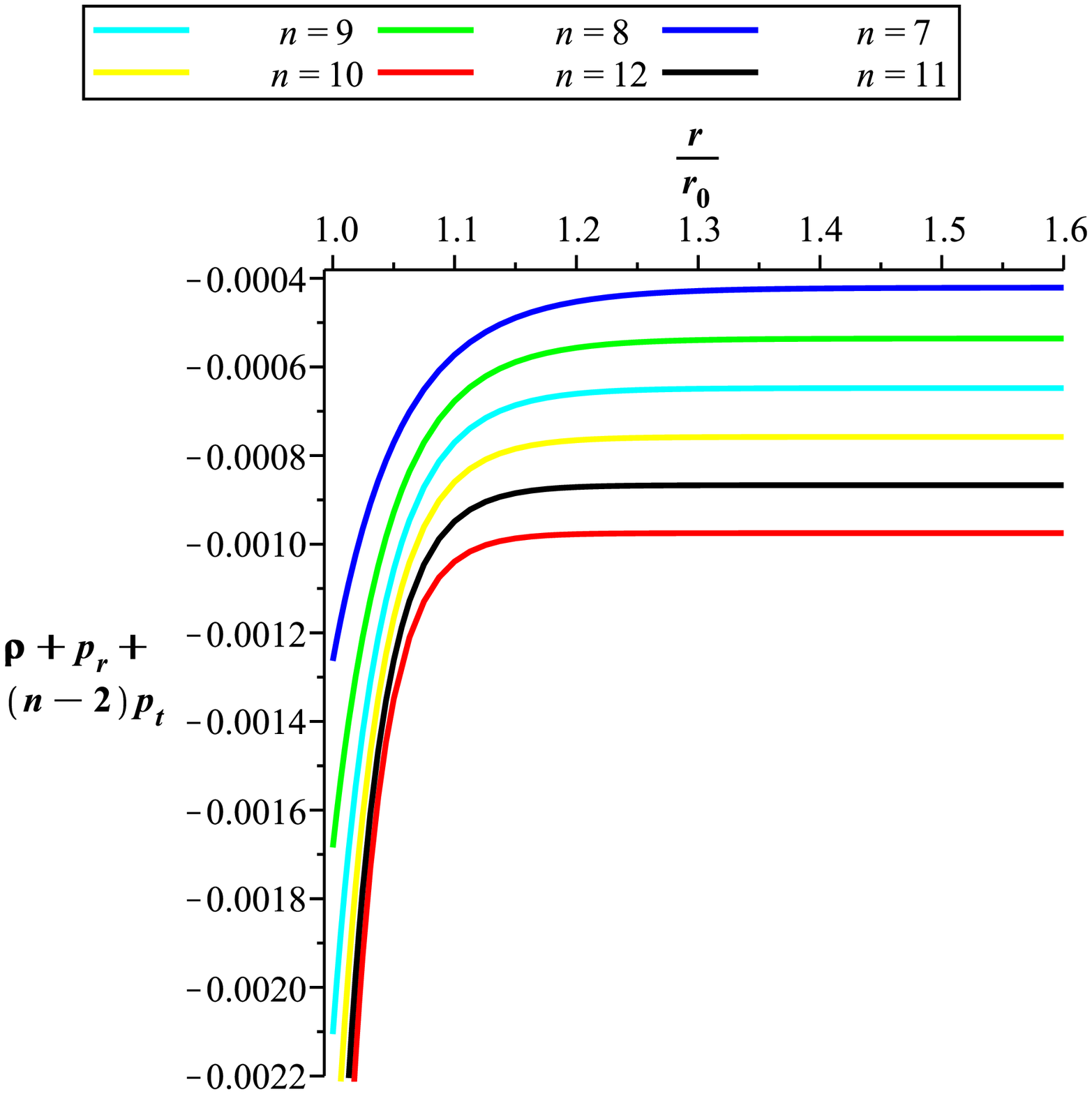}
	\end{tabular}
	\caption{\small{Plot to study $\rho + p_{t}$ and $\rho + p_{r} +(n-2)p_{t}$ taking $r_0=0.9$, $\alpha_2=-0.81$, $B_g=0.0001052631579$ km$^{-2}$ for different value of $n$.} }\label{EC2}
\end{figure*}

It may be noted from Fig. \ref{EC1} and left plot of Fig. \ref{EC2} that NEC and WEC are satisfied by the constituent matter of the WH. From Table \ref{tbl-2}, it is noted that NEC and WEC are satisfied for $0<r_0<1$ and  $-1<\alpha_2<0$. These two constraints are physically acceptable. 

\section{Concluding Remarks}
Several studies \cite{dehghani,mehdizadeh2012} reported the existence of WH geometry without being supported by exotic matter under Lovelock gravity. Being motivated by those results, we explored a WH constituted of quark matter obeying MIT Bag equation of state under third order Lovelock gravity. We take a constant redshift which is compatible with the necessary conditions for the existence of a WH. A specific relation between the second and third order Lovelock coefficients ($\alpha_2$ and $\alpha_3$ respectively) is considered, i.e., $\alpha_3 = \left( \frac{\alpha_2}{2} \right)^2$. As can be noted from this relation, only positive values for $\alpha_3$ are possible. More specifically, we get WH solution supported by quark matter for $0 < r_o < 1$ and $-1 < \alpha_2 < 0$, as the constituent matter satisfy NEC as well as WEC in this range of parameter values.
In this connection we would like to mention that the above result implies that for any observer, the energy density is non-negative \cite{Hawking} in this range of values of parameters. This is in confirmation with the physical nature of quark matter. However, the matter violates SEC everywhere near the throat of the WH. It is important to note that violation of SEC is very common even in classical physics [\cite{Hochberg99}, \cite{Molina99}]. Based on observational data, Visser \cite{Visser97} confirmed the violation of SEC during the evolution of the universe at sometime since the epoch of galaxy formation. It was pointed out by Hawking and Ellis \cite{Hawking} that violation of SEC may take place due to a large negative pressure. A massive scalar field may possibly violate the SEC \cite{Tipler78}. However, a prescription has been provided by Biswas et al. \cite{Biswas2020} pointing out that this type of violation of the SEC, especially for a scalar field with a positive potential and any cosmological inflationary process  \cite{Hawking}, can be easily overcome under an alternative theory of gravity since violation of SEC will cause of violation to the classical regime of GTR. They \cite{Biswas2020} have studied anisotropic spherically symmetric strange star under the background of $f(R,T)$ gravity and shown by exhibiting their result for the strange star candidate PSRJ 1614–2230 that there is no violation of SEC. Therefore, it seems that Lovelock gravity does serves this purpose adequately.
In the present investigation, to make the Einstein field equations simpler we have made a choice of $\Phi = constant$. However, it immediately raises the following issue: to what extent this choice is physically acceptable? So, one can execute the present work with a $\Phi \neq constant$ and explore the possibility to solve the Einstein equations, either analytically or numerically, for getting physically viable solutions.

\section*{Acknowledgments}
KC, FR and SR acknowledge the support from the authority of Inter-University Centre for Astronomy and Astrophysics, Pune, India by providing them Visiting Associateship under which a part of this work was carried out. AA also thanks the authority of IUCAA, Pune, India.

\end{document}